\documentclass[%
 reprint,
 amsmath,amssymb,
 aps,
]{revtex4-1}

\usepackage{graphicx}
\usepackage{xcolor}
\usepackage{dcolumn}
\usepackage{bm}
\usepackage[bottom]{footmisc}

\begin{document}

\preprint{APS/123-QED}

\title{Quantum Turbulent Structure in Light}

\author{Samuel N. Alperin}
\altaffiliation{Now at the Department of Applied Mathematics and Theoretical Physics, University of Cambridge, Cambridge CB3 0WA, UK\\Author Email: alperin@maths.cam.ac.uk}
\affiliation{Department of Physics \& Astronomy, University of Denver, Denver, Colorado 80210, USA}

\author{Abigail L. Grotelueschen}
\affiliation{Department of Physics \& Astronomy, University of Denver, Denver, Colorado 80210, USA}

\author{Mark E. Siemens}
\altaffiliation{Author Email: mark.siemens@du.edu}
 \affiliation{Department of Physics \& Astronomy, University of Denver, Denver, Colorado 80210, USA}

\begin{abstract} 
The infinite superpositions of random plane waves are known to be threaded with vortex line singularities which form complicated tangles and obey strict topological rules. We observe that within these structures a timelike axis appears to emerge with which we can define vortex velocities in a useful way: with both numerical simulations and optical experiments, we show that the statistics of these velocities match those of turbulent quantum fluids such as superfluid helium and atomic Bose-Einstein condensates. These statistics are shown to be independent of system scale. These results raise deep questions about the general nature of quantum chaos and the role of nonlinearity in the structure of turbulence.

\end{abstract}

\maketitle

Despite far predating nearly all other open problems in both mathematics and physics, a complete theory of fluid turbulence remains elusive, and represents one of the greatest open problems of either discipline \cite{Eames702}. Though the problem of turbulence in classical fluids was treated with modern sophistication by Kolmogorov in the decade before \cite{kolmogorov1941local,kolmogorov1941degeneration}, it was not until 1955 that Feynman, who called turbulence the most important unsolved problem of classical physics \cite{feynman2011feynman}, gave the problem a truly modern flavor by merging it with quantum mechanics \cite{FEYNMAN195517}. Since then a great deal of effort has gone into the study of the properties of turbulence in quantum fluids, not only as a way to better understand the dynamics of a general class of complicated quantum systems, but also as a way to understand the general problem of fluid turbulence, with quantum turbulence being studied by some as ``the `skeleton' of ordinary turbulence" \cite{barenghi2016regimes}. With the more recent advance of technologies based on quantum fluidic systems, such as in quantum simulation and computation \cite{berloff2017realizing}, the understanding of quantum turbulence has left the realm of pure theory: a fundamental understanding of quantum turbulence has become critical to the continued development of relevant technologies. Many important systems behave as quantum fluids; the turbulent dynamics of superfluid Helium, Bose Einstein condensates, and polariton condensates are all highly significant areas of current research \cite{sasaki2010benard,kobayashi2007quantum,carusotto2013quantum,nardin2011hydrodynamic,amo2011polariton,hivet2012half}.

Unlike a classical fluid, a quantum fluid is spatially coherent over macroscopic distances. This long range order necessitates a singly defined phase over the field, a constraint which forces the quantization of phase vortices into discrete topological charges. The nature of classical and quantum turbulence is in this way fundamentally different: while turbulence in classical fluids is characterized by structure spanning many length scales, turbulent quantum fluids have a characteristic length scale imparted by vortex quantization. Still, it has only recently been shown that there exists a fundamental difference in the measurable statistical behavior of quantum and classical fluids: the velocity statistics of turbulent quantum fluids differ dramatically from those of turbulent classical fluids \cite{paoletti2008velocity,white2010nonclassical,numasato2010direct,baggaley2011vortex,baggaley2012coherent}. While the velocity statistics of classically turbulent systems show near-Gaussian probability distributions \cite{gotoh2002velocity,noullez1997transverse}, the velocity distributions of quantum turbulence are heavy-tailed \cite{paoletti2008velocity,white2010nonclassical,numasato2010direct,baggaley2011vortex,baggaley2012coherent}. To our knowledge this stands as the only quantitative measure which definitively separates quantum and classical turbulence in the inertial subrange.

While the quantum fluids discussed so far have represented exotic states of condensed matter systems, we suggest that any spatially coherent wave should obey the same fundamental behavior; in this Letter we analogize laser light to a macroscopic, room temperature quantum fluid which supports turbulent structure. Laser light clearly satisfies the requirement of macroscopic coherence, and previous work has shown that the similarity in governing equations can explain dynamical similarities of few vortex systems \cite{rozas1997experimental}. However, it is less obvious that light can support turbulent structure in free space, in which there are no nonlinear interactions. The pioneering work of Nye, Berry and Dennis has revealed qualitative links between random optical fields and superfluid turbulence \cite{nye1974dislocations,berry2000phase,o2009topology}, but others have argued that in a linear system the vortices cannot drive the dynamics of the field and thus the only similarity is appearance \cite{pismen1999vortices}. This has resulted in a widespread assumption that quantum fluidic behavior requires many body interactions \cite{carusotto2013quantum}.

In one sense, a turbulent state is one which is characterized by disorder. If we think of isotropically turbulent quantum fluids found in condensed matter systems as maximally disordered solutions to their governing equations (the nonlinear Schr\"odinger equations), then we should expect that if we are to find structure reminiscent of isotropic quantum turbulence in a linear system, we should look at the maximally disordered solutions of those governing equations. In the case of light, the maximally disordered solutions to the Helmholtz equation are those which can be decomposed into a uniform distribution of random plane waves. This so called \textit{random wave} system is well studied in the context of quantum chaos, modeling the natural excitations of classically chaotic billiard tables \cite{de2017persistence,heller1984bound}, and is also known to be a good model of monochromatic laser scatter \cite{baranova1981dislocations}. 

We begin with the construction of the random wave. Although it is trivial to define a single plane wave with two spatial dimensions and one time (or propagation) axis, as more plane waves of different orientaions are added, the phase front ceases to be confined to one well defined xy-plane. As the number of random plane waves approaches infinity, the phase front of the resulting superposition fills 3-dimensional space. It is thus nontrivial to claim that the z-axis might act as a time axis in any representation of the system. However, we find that although the time axis becomes ill defined from the perspective of constituent plane waves, the propagation axis of the random wave is timelike when viewed from the perspective of emergent vortex behavior.

\begin{figure}[!ht]
\centering
\includegraphics[width=\columnwidth]{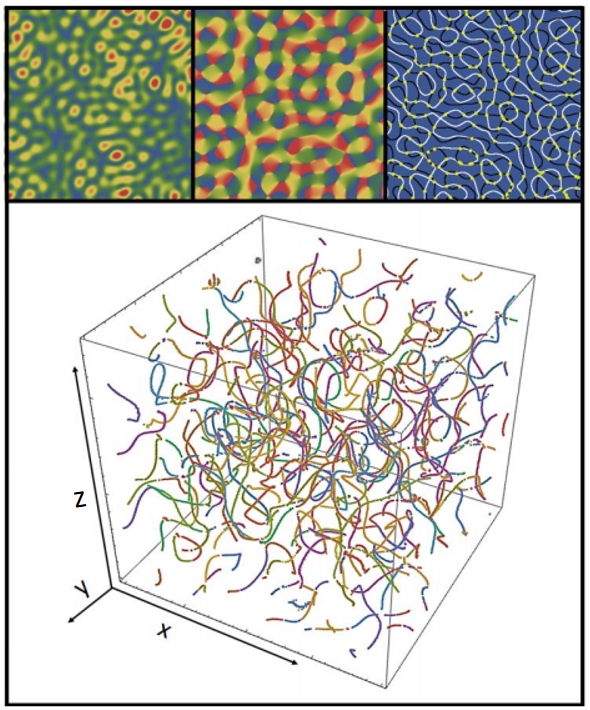}
 \caption{The transverse amplitude (top left) and phase (top middle) of a typical random wave. The top right  panel shows the zeros of the real (white) and imaginary (black) parts of the wave represented in the top two panels, the intersections of which occur at phase singularities, which are marked in yellow. These singularities can be tracked over the propagation of the wave, the paths of which are shown in the bottom panel. That cell is oriented such that the $z$ axis is vertical. Each vortex line is marked in a different color, highlighting a complicated, tangle-like structure. The form of tangled vortex lines shown here appears qualitatively identical to similar plots of vortex lines in condensed matter systems \cite{baggaley2012coherent}.}
  \label{fig1}
\end{figure}

The top of Fig. \ref{fig1} shows the numerically simulated amplitude and phase of a transverse slice of a typical random wave. The lower panel of that figure shows the tangled paths of phase singularities. We define the transverse velocity of each vortex $i$ as the rate of change of vortex position $r_i$ as a function of $z$. 

Random waves are simulated by superposing a finite number of uniformly parameterized plane waves. To propagate the random wave, each plane wave in the superposition is evolved in time. Over this time evolution, the phase singularities of the random wave move about the $xy$-plane, and are created and annihilated in pairs of opposite charge, seemingly at random. However each charge is topologically stable, and because of the constraint of a continuously well defined phase there is more structure than appears, as will become clear shortly.

\begin{figure}[!t]
\centering
 \includegraphics[width=\columnwidth]{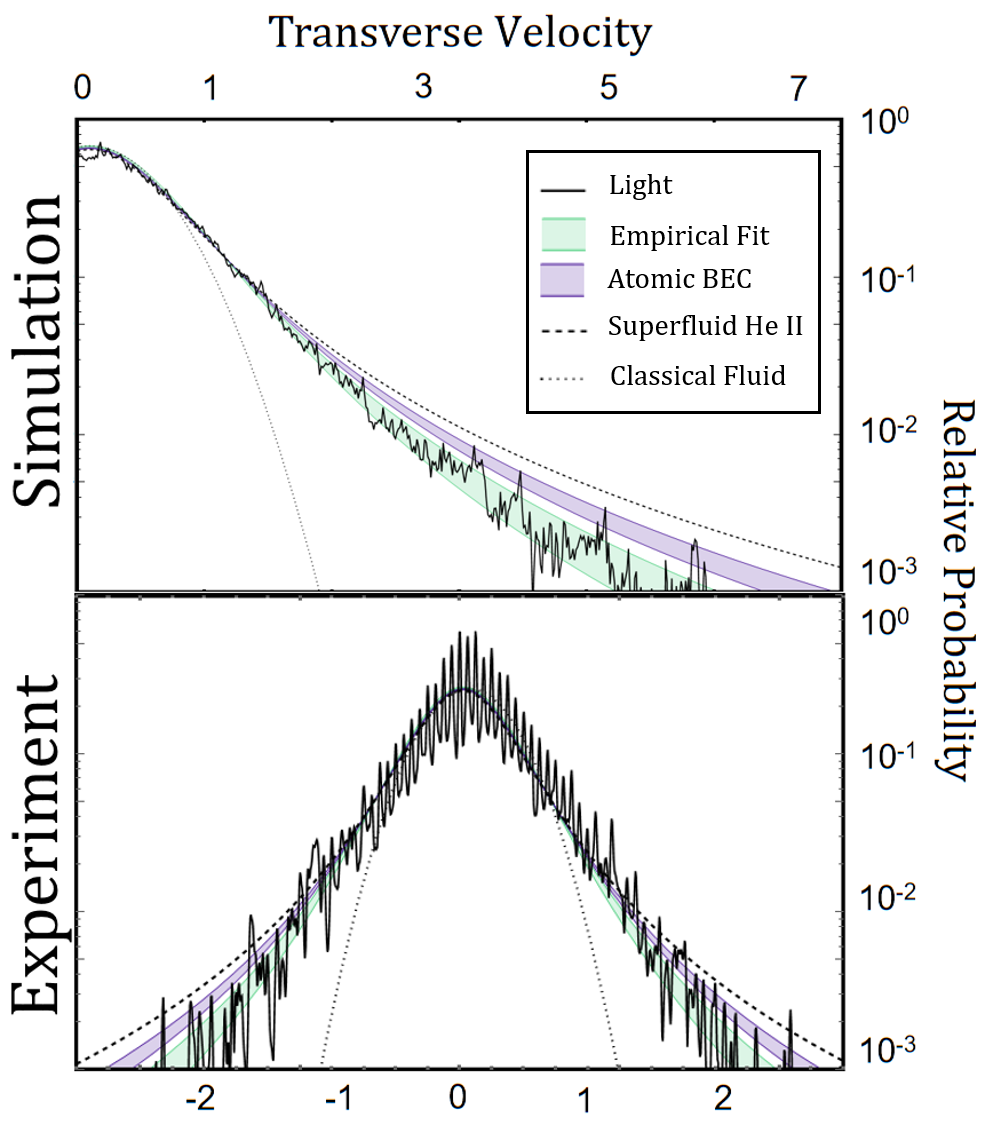}
 \caption{The numerically simulated (top) and experimentally measured (bottom) probability distributions of transverse vortex velocities in random optical scatter (black), plotted on a logarithmic scale showing relative probabilities over three orders of magnitude. Units of velocity are nondimensionalized by scaling by the variance of the distributions. Physically, velocities represent the relative transverse rate of change of vortex position as a function of $z$.  Also plotted are curves consistent with the vortex velocity distributions of the classical fluid (dotted), the atomic BEC (purple band), and superfluid helium (dashed) \cite{paoletti2008velocity,white2010nonclassical}. We fit the superstatistical q-distribution to our results, finding q for our results falls between about 1.4 and 1.5 (green bands in Fig.\ref{fig2}). This corresponds to power-law tails between 3.8 and 4.6. The same curves are shown in the plots of simulated and experimentally measured data, up to rescaling by the variances of each distribution.
 }
  \label{fig2}
\end{figure}

We record the distribution of vortex velocities in our simulated random waves, the results of which are shown at the top of Fig. \ref{fig2}.  This shows that the vortices in linear random waves do indeed obey the characteristically heavy tailed velocity statistics of quantum turbulence. That figure also shows fits to curves consistent with the statistics of classical fluids (dotted), atomic Bose-Einstein condensates (purple band) and superfluid Helium II (dashed). In most other works on quantum turbulent velocity statistics these curves are fit the form of $a/x^b$ for a scaling constant $a$ and tail defined by $b$, which match well at the tails of measured distributions but which are divergent and unphysical near the center \cite{paoletti2008velocity,white2010nonclassical,numasato2010direct,baggaley2011vortex,baggaley2012coherent}. As a more physical model we fit to the superstatistical model of Beck and Miah \cite{beck2013statistics}, which has been shown to match the velocity statistics of quantum fluids in both the low and high velocity regimes.

Fig. \ref{fig2} shows, our optical results closely match the form of the heavy tailed velocity distributions consistent with experiments and simulations of various quantum fluids, and are inconsistent with the statistics of classical fluids, which are Gaussian (dotted) \cite{gotoh2002velocity,noullez1997transverse}. Our results reveal an apparent hierarchy of statistical distributions, with fits to our data having power-law tails between $1/x^{3.8}$ and $1/x^{4.6}$, while the tails of atomic Bose-Einstein condensates have been shown to be between $1/x^{3.3}$ and $1/x^{3.6}$, and those of superfluid helium fitting well to $1/x^3$. This appear to follow from superfluid helium having a higher interaction strength than the atomic Bose-Einstein condensate, and from light having no interactions at all. Previous theories have suggested that the velocity distribution seen in superfluid Helium is the fundamental, topologically mandated distribution \cite{paoletti2008velocity,white2010nonclassical}, but our result suggests that the $1/x^3$ tails associated with superfluid helium are the combined result of topological constraints and other phenomena, the statistical effects of which are yet to be understood.

Random scatter composed of a uniform distribution of plane waves makes for a random wave structure with sub-wavelength nearest neighbor vortex distances \cite{Berry2001}. Although it is easy for our numerical simulations to probe this scale, this fine of structure is extremely difficult to observe experimentally, requiring a complicated and specialized near-field imaging apparatus to measure directly. However, because the only constraint on the movement of vortices is topological, we postulate that the statistical behavior of the quantum fluid represents a topological phase of fluid flow which must be invariant under the scale-changing spatial filtration operation inherent to imaging with a simple lens.

Repeating our numerical simulations with different restrictions on the allowed range of polar angles of constituent plane waves, we observe the same characteristic statistics. This implies that the imaging of a random wave with a decreasing numerical aperture results in the observation of the same dynamical vortex structure -- only with increasing characteristic length scales. Thus the form of the velocity distribution is invariant on the low-pass spatial filtration inherent to the optical imaging of high frequency structure up to a scaling constant. This allows us to implement a simple experimental confirmation of our numerical results.

We therefore proceed to discuss an experimental, macroscopic observation of the vortex velocity distribution of laser scatter. We scatter a collimated Helium-Neon (633nm) laser beam with several layers of Scotch tape, and image the amplitude and phase (see Methods) of the the scatter at increasing distances from the tape. A true random wave has no spatial confinement, and thus has no divergence. To approximate the lack of divergence with a scattered Gaussian beam, we image the scattered mode near the plane of the scatterer, where the mode is at its waist.  

We implement the phase shifting holography technique of Yamaguchi \cite{Yamaguchi1997} to image both the amplitude and complete phase structure of the physical random field over a length of its propagation. Physically, this is simply a two arm interferometer in which one arm can be shifted in length by quarter wavelengths, and in which the sample
is shifted along the propagation axis in steps of 0.05mm. Our imaging objective has a numerical aperture of 0.75. We analyze the resulting 3D complex image as in the numerical simulations.

Marking vortex paths as in our numerical experiments, a distribution of transverse velocities is accumulated. The results, shown at the botton of Fig. \ref{fig2}, reveal that macroscopically observed laser scatter does in fact obey highly non-Gaussian velocity statistics near the surface of the scatterer, consistent with quantum turbulence. As suggested previously, the numerical aperture of our experiment is limited, and consistent with our simulations we see the same behavior as with an ideal near-field imaging apparatus, but at significantly larger length scales. Whereas the ideal optical random wave exhibits inter-vortex lengths on the scale of hundreds of nanometers, we observe inter-vortex lengths on the scale of tens of micrometers, and yet see the exact same characteristic functional form describing the velocity statistics; the velocities in the two systems differ by four orders of magnitude. Continuing to treat this system as one of a 2D sea of point particles which evolve in time, this observed nonreliance of the quantum statistics on interparticle distance tells us that in such systems the effective interactions between vortex singularities do not depend on the distance between them.

It can therefore be said that with topological constraints alone, a completely linear system can be forced to exhibit the emergence of structure typically associated with nonlinear interactions; in condensed matter systems it is understood that it is the nonlinear, particle-like attractions and repulsions of topological charges which create the high velocity creation/annihilation events that cause the heavy-tailed statistics characteristic of quantum turbulence. That these statistics appear in a linear medium is surprising, and raises important questions, as it is not clear what it means for a linear system to possess the characteristic structure of turbulence.

We note that the details of topologically mediated particle-like interactions between vortices in quantum fluids are themselves not yet fully understood. What is clear from our results, however, is that their underlying physics is surprisingly universal, being observed in very different physical systems governed by very different equations.

More generally, we have provided evidence that ordinary laser light can be thought of as a quantum fluid in a useful way, demonstrating that classical optics and quantum fluid dynamics are intimately tied. This raises the question of what other quantum fluid dynamical behavior can be understood with linear wave mechanics. While in this work insight was gained into the fluid dynamics of condensed matter systems with optical analogues, we expect this type of analysis to be fruitful in the other direction as well. In particular, we expect that the structure of vortex nucleations, such as is seen in the fluid dynamical von K\'arm\'an effect \cite{kwon2016observation}, will soon inspire the study of new optical phenomena, and bring insight to visually similar and perhaps dynamically analogous optical structures such as the fractional vortex Hilbert's hotel \cite{Gbur2016}. In short, it seems that members of both fields have much more to learn from each other than previously realized.

  We thank Geoffrey Diederich for helpful experimental advice, and thank Mark Lusk for enlightening discussions and for commenting in detail on our manuscript. The authors acknowledge financial support from the National Science Foundation (1509733,1553905).

\bibliography{lib}

\begin{thebibliography}{33}%
\makeatletter
\providecommand \@ifxundefined [1]{%
 \@ifx{#1\undefined}
}%
\providecommand \@ifnum [1]{%
 \ifnum #1\expandafter \@firstoftwo
 \else \expandafter \@secondoftwo
 \fi
}%
\providecommand \@ifx [1]{%
 \ifx #1\expandafter \@firstoftwo
 \else \expandafter \@secondoftwo
 \fi
}%
\providecommand \natexlab [1]{#1}%
\providecommand \enquote  [1]{``#1''}%
\providecommand \bibnamefont  [1]{#1}%
\providecommand \bibfnamefont [1]{#1}%
\providecommand \citenamefont [1]{#1}%
\providecommand \href@noop [0]{\@secondoftwo}%
\providecommand \href [0]{\begingroup \@sanitize@url \@href}%
\providecommand \@href[1]{\@@startlink{#1}\@@href}%
\providecommand \@@href[1]{\endgroup#1\@@endlink}%
\providecommand \@sanitize@url [0]{\catcode `\\12\catcode `\$12\catcode
  `\&12\catcode `\#12\catcode `\^12\catcode `\_12\catcode `\%12\relax}%
\providecommand \@@startlink[1]{}%
\providecommand \@@endlink[0]{}%
\providecommand \url  [0]{\begingroup\@sanitize@url \@url }%
\providecommand \@url [1]{\endgroup\@href {#1}{\urlprefix }}%
\providecommand \urlprefix  [0]{URL }%
\providecommand \Eprint [0]{\href }%
\providecommand \doibase [0]{http://dx.doi.org/}%
\providecommand \selectlanguage [0]{\@gobble}%
\providecommand \bibinfo  [0]{\@secondoftwo}%
\providecommand \bibfield  [0]{\@secondoftwo}%
\providecommand \translation [1]{[#1]}%
\providecommand \BibitemOpen [0]{}%
\providecommand \bibitemStop [0]{}%
\providecommand \bibitemNoStop [0]{.\EOS\space}%
\providecommand \EOS [0]{\spacefactor3000\relax}%
\providecommand \BibitemShut  [1]{\csname bibitem#1\endcsname}%
\let\auto@bib@innerbib\@empty
\bibitem [{\citenamefont {Eames}\ and\ \citenamefont {Flor}(2011)}]{Eames702}%
  \BibitemOpen
  \bibfield  {author} {\bibinfo {author} {\bibfnamefont {I.}~\bibnamefont
  {Eames}}\ and\ \bibinfo {author} {\bibfnamefont {J.~B.}\ \bibnamefont
  {Flor}},\ }\href {\doibase 10.1098/rsta.2010.0332} {\bibfield  {journal}
  {\bibinfo  {journal} {Philosophical Transactions of the Royal Society of
  London A: Mathematical, Physical and Engineering Sciences}\ }\textbf
  {\bibinfo {volume} {369}},\ \bibinfo {pages} {702} (\bibinfo {year}
  {2011})}\BibitemShut {NoStop}%
\bibitem [{\citenamefont
  {Kolmogorov}(1941{\natexlab{a}})}]{kolmogorov1941local}%
  \BibitemOpen
  \bibfield  {author} {\bibinfo {author} {\bibfnamefont {A.~N.}\ \bibnamefont
  {Kolmogorov}},\ }in\ \href@noop {} {\emph {\bibinfo {booktitle} {Dokl. Akad.
  Nauk SSSR}}},\ Vol.~\bibinfo {volume} {30}\ (\bibinfo {year} {1941})\ pp.\
  \bibinfo {pages} {299--303}\BibitemShut {NoStop}%
\bibitem [{\citenamefont
  {Kolmogorov}(1941{\natexlab{b}})}]{kolmogorov1941degeneration}%
  \BibitemOpen
  \bibfield  {author} {\bibinfo {author} {\bibfnamefont {A.~N.}\ \bibnamefont
  {Kolmogorov}},\ }in\ \href@noop {} {\emph {\bibinfo {booktitle} {Dokl. Akad.
  Nauk SSSR}}},\ Vol.~\bibinfo {volume} {31}\ (\bibinfo {year} {1941})\ pp.\
  \bibinfo {pages} {319--323}\BibitemShut {NoStop}%
\bibitem [{\citenamefont {Feynman}\ \emph {et~al.}(2011)\citenamefont
  {Feynman}, \citenamefont {Leighton},\ and\ \citenamefont
  {Sands}}]{feynman2011feynman}%
  \BibitemOpen
  \bibfield  {author} {\bibinfo {author} {\bibfnamefont {R.~P.}\ \bibnamefont
  {Feynman}}, \bibinfo {author} {\bibfnamefont {R.~B.}\ \bibnamefont
  {Leighton}}, \ and\ \bibinfo {author} {\bibfnamefont {M.}~\bibnamefont
  {Sands}},\ }\href@noop {} {\emph {\bibinfo {title} {The Feynman lectures on
  physics, Vol. I: The new millennium edition: mainly mechanics, radiation, and
  heat}}},\ Vol.~\bibinfo {volume} {1}\ (\bibinfo  {publisher} {Basic books},\
  \bibinfo {year} {2011})\BibitemShut {NoStop}%
\bibitem [{\citenamefont {Feynman}(1955)}]{FEYNMAN195517}%
  \BibitemOpen
  \bibfield  {author} {\bibinfo {author} {\bibfnamefont {R.}~\bibnamefont
  {Feynman}},\ }\href {\doibase https://doi.org/10.1016/S0079-6417(08)60077-3}
  {\bibfield  {journal} {\bibinfo  {journal} {Progress in Low Temperature
  Physics}\ }\textbf {\bibinfo {volume} {1}},\ \bibinfo {pages} {17 } (\bibinfo
  {year} {1955})}\BibitemShut {NoStop}%
\bibitem [{\citenamefont {Barenghi}\ \emph {et~al.}(2016)\citenamefont
  {Barenghi}, \citenamefont {Sergeev},\ and\ \citenamefont
  {Baggaley}}]{barenghi2016regimes}%
  \BibitemOpen
  \bibfield  {author} {\bibinfo {author} {\bibfnamefont {C.}~\bibnamefont
  {Barenghi}}, \bibinfo {author} {\bibfnamefont {Y.}~\bibnamefont {Sergeev}}, \
  and\ \bibinfo {author} {\bibfnamefont {A.}~\bibnamefont {Baggaley}},\
  }\href@noop {} {\bibfield  {journal} {\bibinfo  {journal} {Scientific
  Reports}\ }\textbf {\bibinfo {volume} {6}},\ \bibinfo {pages} {35701}
  (\bibinfo {year} {2016})}\BibitemShut {NoStop}%
\bibitem [{\citenamefont {Berloff}\ \emph {et~al.}(2017)\citenamefont
  {Berloff}, \citenamefont {Silva}, \citenamefont {Kalinin}, \citenamefont
  {Askitopoulos}, \citenamefont {T{\"o}pfer}, \citenamefont {Cilibrizzi},
  \citenamefont {Langbein},\ and\ \citenamefont
  {Lagoudakis}}]{berloff2017realizing}%
  \BibitemOpen
  \bibfield  {author} {\bibinfo {author} {\bibfnamefont {N.~G.}\ \bibnamefont
  {Berloff}}, \bibinfo {author} {\bibfnamefont {M.}~\bibnamefont {Silva}},
  \bibinfo {author} {\bibfnamefont {K.}~\bibnamefont {Kalinin}}, \bibinfo
  {author} {\bibfnamefont {A.}~\bibnamefont {Askitopoulos}}, \bibinfo {author}
  {\bibfnamefont {J.~D.}\ \bibnamefont {T{\"o}pfer}}, \bibinfo {author}
  {\bibfnamefont {P.}~\bibnamefont {Cilibrizzi}}, \bibinfo {author}
  {\bibfnamefont {W.}~\bibnamefont {Langbein}}, \ and\ \bibinfo {author}
  {\bibfnamefont {P.~G.}\ \bibnamefont {Lagoudakis}},\ }\href@noop {}
  {\bibfield  {journal} {\bibinfo  {journal} {Nature materials}\ }\textbf
  {\bibinfo {volume} {16}},\ \bibinfo {pages} {1120} (\bibinfo {year}
  {2017})}\BibitemShut {NoStop}%
\bibitem [{\citenamefont {Sasaki}\ \emph {et~al.}(2010)\citenamefont {Sasaki},
  \citenamefont {Suzuki},\ and\ \citenamefont {Saito}}]{sasaki2010benard}%
  \BibitemOpen
  \bibfield  {author} {\bibinfo {author} {\bibfnamefont {K.}~\bibnamefont
  {Sasaki}}, \bibinfo {author} {\bibfnamefont {N.}~\bibnamefont {Suzuki}}, \
  and\ \bibinfo {author} {\bibfnamefont {H.}~\bibnamefont {Saito}},\
  }\href@noop {} {\bibfield  {journal} {\bibinfo  {journal} {Physical review
  letters}\ }\textbf {\bibinfo {volume} {104}},\ \bibinfo {pages} {150404}
  (\bibinfo {year} {2010})}\BibitemShut {NoStop}%
\bibitem [{\citenamefont {Kobayashi}\ and\ \citenamefont
  {Tsubota}(2007)}]{kobayashi2007quantum}%
  \BibitemOpen
  \bibfield  {author} {\bibinfo {author} {\bibfnamefont {M.}~\bibnamefont
  {Kobayashi}}\ and\ \bibinfo {author} {\bibfnamefont {M.}~\bibnamefont
  {Tsubota}},\ }\href@noop {} {\bibfield  {journal} {\bibinfo  {journal}
  {Physical Review A}\ }\textbf {\bibinfo {volume} {76}},\ \bibinfo {pages}
  {045603} (\bibinfo {year} {2007})}\BibitemShut {NoStop}%
\bibitem [{\citenamefont {Carusotto}\ and\ \citenamefont
  {Ciuti}(2013)}]{carusotto2013quantum}%
  \BibitemOpen
  \bibfield  {author} {\bibinfo {author} {\bibfnamefont {I.}~\bibnamefont
  {Carusotto}}\ and\ \bibinfo {author} {\bibfnamefont {C.}~\bibnamefont
  {Ciuti}},\ }\href@noop {} {\bibfield  {journal} {\bibinfo  {journal} {Reviews
  of Modern Physics}\ }\textbf {\bibinfo {volume} {85}},\ \bibinfo {pages}
  {299} (\bibinfo {year} {2013})}\BibitemShut {NoStop}%
\bibitem [{\citenamefont {Nardin}\ \emph {et~al.}(2011)\citenamefont {Nardin},
  \citenamefont {Grosso}, \citenamefont {L{\'e}ger}, \citenamefont {Piȩtka},
  \citenamefont {Morier-Genoud},\ and\ \citenamefont
  {Deveaud-Pl{\'e}dran}}]{nardin2011hydrodynamic}%
  \BibitemOpen
  \bibfield  {author} {\bibinfo {author} {\bibfnamefont {G.}~\bibnamefont
  {Nardin}}, \bibinfo {author} {\bibfnamefont {G.}~\bibnamefont {Grosso}},
  \bibinfo {author} {\bibfnamefont {Y.}~\bibnamefont {L{\'e}ger}}, \bibinfo
  {author} {\bibfnamefont {B.}~\bibnamefont {Piȩtka}}, \bibinfo {author}
  {\bibfnamefont {F.}~\bibnamefont {Morier-Genoud}}, \ and\ \bibinfo {author}
  {\bibfnamefont {B.}~\bibnamefont {Deveaud-Pl{\'e}dran}},\ }\href@noop {}
  {\bibfield  {journal} {\bibinfo  {journal} {Nature Physics}\ }\textbf
  {\bibinfo {volume} {7}},\ \bibinfo {pages} {635} (\bibinfo {year}
  {2011})}\BibitemShut {NoStop}%
\bibitem [{\citenamefont {Amo}\ \emph {et~al.}(2011)\citenamefont {Amo},
  \citenamefont {Pigeon}, \citenamefont {Sanvitto}, \citenamefont {Sala},
  \citenamefont {Hivet}, \citenamefont {Carusotto}, \citenamefont {Pisanello},
  \citenamefont {Lem{\'e}nager}, \citenamefont {Houdr{\'e}}, \citenamefont
  {Giacobino} \emph {et~al.}}]{amo2011polariton}%
  \BibitemOpen
  \bibfield  {author} {\bibinfo {author} {\bibfnamefont {A.}~\bibnamefont
  {Amo}}, \bibinfo {author} {\bibfnamefont {S.}~\bibnamefont {Pigeon}},
  \bibinfo {author} {\bibfnamefont {D.}~\bibnamefont {Sanvitto}}, \bibinfo
  {author} {\bibfnamefont {V.}~\bibnamefont {Sala}}, \bibinfo {author}
  {\bibfnamefont {R.}~\bibnamefont {Hivet}}, \bibinfo {author} {\bibfnamefont
  {I.}~\bibnamefont {Carusotto}}, \bibinfo {author} {\bibfnamefont
  {F.}~\bibnamefont {Pisanello}}, \bibinfo {author} {\bibfnamefont
  {G.}~\bibnamefont {Lem{\'e}nager}}, \bibinfo {author} {\bibfnamefont
  {R.}~\bibnamefont {Houdr{\'e}}}, \bibinfo {author} {\bibfnamefont
  {E.}~\bibnamefont {Giacobino}},  \emph {et~al.},\ }\href@noop {} {\bibfield
  {journal} {\bibinfo  {journal} {Science}\ }\textbf {\bibinfo {volume}
  {332}},\ \bibinfo {pages} {1167} (\bibinfo {year} {2011})}\BibitemShut
  {NoStop}%
\bibitem [{\citenamefont {Hivet}\ \emph {et~al.}(2012)\citenamefont {Hivet},
  \citenamefont {Flayac}, \citenamefont {Solnyshkov}, \citenamefont {Tanese},
  \citenamefont {Boulier}, \citenamefont {Andreoli}, \citenamefont {Giacobino},
  \citenamefont {Bloch}, \citenamefont {Bramati}, \citenamefont {Malpuech}
  \emph {et~al.}}]{hivet2012half}%
  \BibitemOpen
  \bibfield  {author} {\bibinfo {author} {\bibfnamefont {R.}~\bibnamefont
  {Hivet}}, \bibinfo {author} {\bibfnamefont {H.}~\bibnamefont {Flayac}},
  \bibinfo {author} {\bibfnamefont {D.}~\bibnamefont {Solnyshkov}}, \bibinfo
  {author} {\bibfnamefont {D.}~\bibnamefont {Tanese}}, \bibinfo {author}
  {\bibfnamefont {T.}~\bibnamefont {Boulier}}, \bibinfo {author} {\bibfnamefont
  {D.}~\bibnamefont {Andreoli}}, \bibinfo {author} {\bibfnamefont
  {E.}~\bibnamefont {Giacobino}}, \bibinfo {author} {\bibfnamefont
  {J.}~\bibnamefont {Bloch}}, \bibinfo {author} {\bibfnamefont
  {A.}~\bibnamefont {Bramati}}, \bibinfo {author} {\bibfnamefont
  {G.}~\bibnamefont {Malpuech}},  \emph {et~al.},\ }\href@noop {} {\bibfield
  {journal} {\bibinfo  {journal} {Nature Physics}\ }\textbf {\bibinfo {volume}
  {8}},\ \bibinfo {pages} {724} (\bibinfo {year} {2012})}\BibitemShut {NoStop}%
\bibitem [{\citenamefont {Paoletti}\ \emph {et~al.}(2008)\citenamefont
  {Paoletti}, \citenamefont {Fisher}, \citenamefont {Sreenivasan},\ and\
  \citenamefont {Lathrop}}]{paoletti2008velocity}%
  \BibitemOpen
  \bibfield  {author} {\bibinfo {author} {\bibfnamefont {M.~S.}\ \bibnamefont
  {Paoletti}}, \bibinfo {author} {\bibfnamefont {M.~E.}\ \bibnamefont
  {Fisher}}, \bibinfo {author} {\bibfnamefont {K.~R.}\ \bibnamefont
  {Sreenivasan}}, \ and\ \bibinfo {author} {\bibfnamefont {D.~P.}\ \bibnamefont
  {Lathrop}},\ }\href@noop {} {\bibfield  {journal} {\bibinfo  {journal}
  {Physical review letters}\ }\textbf {\bibinfo {volume} {101}},\ \bibinfo
  {pages} {154501} (\bibinfo {year} {2008})}\BibitemShut {NoStop}%
\bibitem [{\citenamefont {White}\ \emph {et~al.}(2010)\citenamefont {White},
  \citenamefont {Barenghi}, \citenamefont {Proukakis}, \citenamefont {Youd},\
  and\ \citenamefont {Wacks}}]{white2010nonclassical}%
  \BibitemOpen
  \bibfield  {author} {\bibinfo {author} {\bibfnamefont {A.}~\bibnamefont
  {White}}, \bibinfo {author} {\bibfnamefont {C.}~\bibnamefont {Barenghi}},
  \bibinfo {author} {\bibfnamefont {N.}~\bibnamefont {Proukakis}}, \bibinfo
  {author} {\bibfnamefont {A.}~\bibnamefont {Youd}}, \ and\ \bibinfo {author}
  {\bibfnamefont {D.}~\bibnamefont {Wacks}},\ }\href@noop {} {\bibfield
  {journal} {\bibinfo  {journal} {Physical review letters}\ }\textbf {\bibinfo
  {volume} {104}},\ \bibinfo {pages} {075301} (\bibinfo {year}
  {2010})}\BibitemShut {NoStop}%
\bibitem [{\citenamefont {Numasato}\ \emph {et~al.}(2010)\citenamefont
  {Numasato}, \citenamefont {Tsubota},\ and\ \citenamefont
  {L’vov}}]{numasato2010direct}%
  \BibitemOpen
  \bibfield  {author} {\bibinfo {author} {\bibfnamefont {R.}~\bibnamefont
  {Numasato}}, \bibinfo {author} {\bibfnamefont {M.}~\bibnamefont {Tsubota}}, \
  and\ \bibinfo {author} {\bibfnamefont {V.~S.}\ \bibnamefont {L’vov}},\
  }\href@noop {} {\bibfield  {journal} {\bibinfo  {journal} {Physical Review
  A}\ }\textbf {\bibinfo {volume} {81}},\ \bibinfo {pages} {063630} (\bibinfo
  {year} {2010})}\BibitemShut {NoStop}%
\bibitem [{\citenamefont {Baggaley}\ and\ \citenamefont
  {Barenghi}(2011)}]{baggaley2011vortex}%
  \BibitemOpen
  \bibfield  {author} {\bibinfo {author} {\bibfnamefont {A.~W.}\ \bibnamefont
  {Baggaley}}\ and\ \bibinfo {author} {\bibfnamefont {C.~F.}\ \bibnamefont
  {Barenghi}},\ }\href@noop {} {\bibfield  {journal} {\bibinfo  {journal}
  {Physical Review B}\ }\textbf {\bibinfo {volume} {84}},\ \bibinfo {pages}
  {020504} (\bibinfo {year} {2011})}\BibitemShut {NoStop}%
\bibitem [{\citenamefont {Baggaley}\ \emph {et~al.}(2012)\citenamefont
  {Baggaley}, \citenamefont {Barenghi}, \citenamefont {Shukurov},\ and\
  \citenamefont {Sergeev}}]{baggaley2012coherent}%
  \BibitemOpen
  \bibfield  {author} {\bibinfo {author} {\bibfnamefont {A.~W.}\ \bibnamefont
  {Baggaley}}, \bibinfo {author} {\bibfnamefont {C.~F.}\ \bibnamefont
  {Barenghi}}, \bibinfo {author} {\bibfnamefont {A.}~\bibnamefont {Shukurov}},
  \ and\ \bibinfo {author} {\bibfnamefont {Y.~A.}\ \bibnamefont {Sergeev}},\
  }\href@noop {} {\bibfield  {journal} {\bibinfo  {journal} {EPL (Europhysics
  Letters)}\ }\textbf {\bibinfo {volume} {98}},\ \bibinfo {pages} {26002}
  (\bibinfo {year} {2012})}\BibitemShut {NoStop}%
\bibitem [{\citenamefont {Gotoh}\ \emph {et~al.}(2002)\citenamefont {Gotoh},
  \citenamefont {Fukayama},\ and\ \citenamefont {Nakano}}]{gotoh2002velocity}%
  \BibitemOpen
  \bibfield  {author} {\bibinfo {author} {\bibfnamefont {T.}~\bibnamefont
  {Gotoh}}, \bibinfo {author} {\bibfnamefont {D.}~\bibnamefont {Fukayama}}, \
  and\ \bibinfo {author} {\bibfnamefont {T.}~\bibnamefont {Nakano}},\
  }\href@noop {} {\bibfield  {journal} {\bibinfo  {journal} {Physics of
  Fluids}\ }\textbf {\bibinfo {volume} {14}},\ \bibinfo {pages} {1065}
  (\bibinfo {year} {2002})}\BibitemShut {NoStop}%
\bibitem [{\citenamefont {Noullez}\ \emph {et~al.}(1997)\citenamefont
  {Noullez}, \citenamefont {Wallace}, \citenamefont {Lempert}, \citenamefont
  {Miles},\ and\ \citenamefont {Frisch}}]{noullez1997transverse}%
  \BibitemOpen
  \bibfield  {author} {\bibinfo {author} {\bibfnamefont {A.}~\bibnamefont
  {Noullez}}, \bibinfo {author} {\bibfnamefont {G.}~\bibnamefont {Wallace}},
  \bibinfo {author} {\bibfnamefont {W.}~\bibnamefont {Lempert}}, \bibinfo
  {author} {\bibfnamefont {R.}~\bibnamefont {Miles}}, \ and\ \bibinfo {author}
  {\bibfnamefont {U.}~\bibnamefont {Frisch}},\ }\href@noop {} {\bibfield
  {journal} {\bibinfo  {journal} {Journal of Fluid Mechanics}\ }\textbf
  {\bibinfo {volume} {339}},\ \bibinfo {pages} {287} (\bibinfo {year}
  {1997})}\BibitemShut {NoStop}%
\bibitem [{\citenamefont {Rozas}\ \emph {et~al.}(1997)\citenamefont {Rozas},
  \citenamefont {Sacks},\ and\ \citenamefont
  {Swartzlander~Jr}}]{rozas1997experimental}%
  \BibitemOpen
  \bibfield  {author} {\bibinfo {author} {\bibfnamefont {D.}~\bibnamefont
  {Rozas}}, \bibinfo {author} {\bibfnamefont {Z.}~\bibnamefont {Sacks}}, \ and\
  \bibinfo {author} {\bibfnamefont {G.}~\bibnamefont {Swartzlander~Jr}},\
  }\href@noop {} {\bibfield  {journal} {\bibinfo  {journal} {Physical review
  letters}\ }\textbf {\bibinfo {volume} {79}},\ \bibinfo {pages} {3399}
  (\bibinfo {year} {1997})}\BibitemShut {NoStop}%
\bibitem [{\citenamefont {Nye}\ and\ \citenamefont
  {Berry}(1974)}]{nye1974dislocations}%
  \BibitemOpen
  \bibfield  {author} {\bibinfo {author} {\bibfnamefont {J.}~\bibnamefont
  {Nye}}\ and\ \bibinfo {author} {\bibfnamefont {M.}~\bibnamefont {Berry}},\
  }\href@noop {} {\bibfield  {journal} {\bibinfo  {journal} {Proc. R. Soc.
  Lond. A}\ }\textbf {\bibinfo {volume} {336}},\ \bibinfo {pages} {165}
  (\bibinfo {year} {1974})}\BibitemShut {NoStop}%
\bibitem [{\citenamefont {Berry}\ and\ \citenamefont
  {Dennis}(2000)}]{berry2000phase}%
  \BibitemOpen
  \bibfield  {author} {\bibinfo {author} {\bibfnamefont {M.}~\bibnamefont
  {Berry}}\ and\ \bibinfo {author} {\bibfnamefont {M.}~\bibnamefont {Dennis}},\
  }\href@noop {} {\bibfield  {journal} {\bibinfo  {journal} {Proc. R. Soc.
  Lond. A}\ }\textbf {\bibinfo {volume} {456}},\ \bibinfo {pages} {2059}
  (\bibinfo {year} {2000})}\BibitemShut {NoStop}%
\bibitem [{\citenamefont {O’Holleran}\ \emph {et~al.}(2009)\citenamefont
  {O’Holleran}, \citenamefont {Dennis},\ and\ \citenamefont
  {Padgett}}]{o2009topology}%
  \BibitemOpen
  \bibfield  {author} {\bibinfo {author} {\bibfnamefont {K.}~\bibnamefont
  {O’Holleran}}, \bibinfo {author} {\bibfnamefont {M.~R.}\ \bibnamefont
  {Dennis}}, \ and\ \bibinfo {author} {\bibfnamefont {M.~J.}\ \bibnamefont
  {Padgett}},\ }\href@noop {} {\bibfield  {journal} {\bibinfo  {journal}
  {Physical review letters}\ }\textbf {\bibinfo {volume} {102}},\ \bibinfo
  {pages} {143902} (\bibinfo {year} {2009})}\BibitemShut {NoStop}%
\bibitem [{\citenamefont {Pismen}(1999)}]{pismen1999vortices}%
  \BibitemOpen
  \bibfield  {author} {\bibinfo {author} {\bibfnamefont {L.~M.}\ \bibnamefont
  {Pismen}},\ }\href@noop {} {\emph {\bibinfo {title} {Vortices in nonlinear
  fields: From liquid crystals to superfluids, from non-equilibrium patterns to
  cosmic strings}}},\ Vol.\ \bibinfo {volume} {100}\ (\bibinfo  {publisher}
  {Oxford University Press},\ \bibinfo {year} {1999})\BibitemShut {NoStop}%
\bibitem [{\citenamefont {De~Angelis}\ \emph {et~al.}(2017)\citenamefont
  {De~Angelis}, \citenamefont {Alpeggiani}, \citenamefont {Di~Falco},\ and\
  \citenamefont {Kuipers}}]{de2017persistence}%
  \BibitemOpen
  \bibfield  {author} {\bibinfo {author} {\bibfnamefont {L.}~\bibnamefont
  {De~Angelis}}, \bibinfo {author} {\bibfnamefont {F.}~\bibnamefont
  {Alpeggiani}}, \bibinfo {author} {\bibfnamefont {A.}~\bibnamefont
  {Di~Falco}}, \ and\ \bibinfo {author} {\bibfnamefont {L.}~\bibnamefont
  {Kuipers}},\ }\href@noop {} {\bibfield  {journal} {\bibinfo  {journal}
  {Physical review letters}\ }\textbf {\bibinfo {volume} {119}},\ \bibinfo
  {pages} {203903} (\bibinfo {year} {2017})}\BibitemShut {NoStop}%
\bibitem [{\citenamefont {Heller}(1984)}]{heller1984bound}%
  \BibitemOpen
  \bibfield  {author} {\bibinfo {author} {\bibfnamefont {E.~J.}\ \bibnamefont
  {Heller}},\ }\href@noop {} {\bibfield  {journal} {\bibinfo  {journal}
  {Physical Review Letters}\ }\textbf {\bibinfo {volume} {53}},\ \bibinfo
  {pages} {1515} (\bibinfo {year} {1984})}\BibitemShut {NoStop}%
\bibitem [{\citenamefont {Baranova}\ \emph {et~al.}(1981)\citenamefont
  {Baranova}, \citenamefont {Zel’Dovich}, \citenamefont {Mamaev},
  \citenamefont {Pilipetskii},\ and\ \citenamefont
  {Shkukov}}]{baranova1981dislocations}%
  \BibitemOpen
  \bibfield  {author} {\bibinfo {author} {\bibfnamefont {N.}~\bibnamefont
  {Baranova}}, \bibinfo {author} {\bibfnamefont {B.~Y.}\ \bibnamefont
  {Zel’Dovich}}, \bibinfo {author} {\bibfnamefont {A.}~\bibnamefont
  {Mamaev}}, \bibinfo {author} {\bibfnamefont {N.}~\bibnamefont {Pilipetskii}},
  \ and\ \bibinfo {author} {\bibfnamefont {V.}~\bibnamefont {Shkukov}},\
  }\href@noop {} {\bibfield  {journal} {\bibinfo  {journal} {Jetp Lett}\
  }\textbf {\bibinfo {volume} {33}},\ \bibinfo {pages} {206} (\bibinfo {year}
  {1981})}\BibitemShut {NoStop}%
\bibitem [{\citenamefont {Beck}\ and\ \citenamefont
  {Miah}(2013)}]{beck2013statistics}%
  \BibitemOpen
  \bibfield  {author} {\bibinfo {author} {\bibfnamefont {C.}~\bibnamefont
  {Beck}}\ and\ \bibinfo {author} {\bibfnamefont {S.}~\bibnamefont {Miah}},\
  }\href@noop {} {\bibfield  {journal} {\bibinfo  {journal} {Physical Review
  E}\ }\textbf {\bibinfo {volume} {87}},\ \bibinfo {pages} {031002} (\bibinfo
  {year} {2013})}\BibitemShut {NoStop}%
\bibitem [{\citenamefont {Berry}\ and\ \citenamefont
  {Dennis}(2001)}]{Berry2001}%
  \BibitemOpen
  \bibfield  {author} {\bibinfo {author} {\bibfnamefont {M.~V.}\ \bibnamefont
  {Berry}}\ and\ \bibinfo {author} {\bibfnamefont {M.~R.}\ \bibnamefont
  {Dennis}},\ }\href {\doibase 10.1088/0305-4470/34/42/311} {\bibfield
  {journal} {\bibinfo  {journal} {J. Phys. A: Math. Gen.}\ }\textbf {\bibinfo
  {volume} {34}},\ \bibinfo {pages} {8877} (\bibinfo {year}
  {2001})}\BibitemShut {NoStop}%
\bibitem [{\citenamefont {Yamaguchi}\ and\ \citenamefont
  {Zhang}(1997)}]{Yamaguchi1997}%
  \BibitemOpen
  \bibfield  {author} {\bibinfo {author} {\bibfnamefont {I.}~\bibnamefont
  {Yamaguchi}}\ and\ \bibinfo {author} {\bibfnamefont {T.}~\bibnamefont
  {Zhang}},\ }\href {\doibase 10.1364/OL.22.001268} {\bibfield  {journal}
  {\bibinfo  {journal} {Optics Letters}\ }\textbf {\bibinfo {volume} {22}},\
  \bibinfo {pages} {1268} (\bibinfo {year} {1997})}\BibitemShut {NoStop}%
\bibitem [{\citenamefont {Kwon}\ \emph {et~al.}(2016)\citenamefont {Kwon},
  \citenamefont {Kim}, \citenamefont {Seo},\ and\ \citenamefont
  {Shin}}]{kwon2016observation}%
  \BibitemOpen
  \bibfield  {author} {\bibinfo {author} {\bibfnamefont {W.~J.}\ \bibnamefont
  {Kwon}}, \bibinfo {author} {\bibfnamefont {J.~H.}\ \bibnamefont {Kim}},
  \bibinfo {author} {\bibfnamefont {S.~W.}\ \bibnamefont {Seo}}, \ and\
  \bibinfo {author} {\bibfnamefont {Y.-i.}\ \bibnamefont {Shin}},\ }\href@noop
  {} {\bibfield  {journal} {\bibinfo  {journal} {Physical review letters}\
  }\textbf {\bibinfo {volume} {117}},\ \bibinfo {pages} {245301} (\bibinfo
  {year} {2016})}\BibitemShut {NoStop}%
\bibitem [{\citenamefont {Gbur}(2016)}]{Gbur2016}%
  \BibitemOpen
  \bibfield  {author} {\bibinfo {author} {\bibfnamefont {G.}~\bibnamefont
  {Gbur}},\ }\href {\doibase 10.1364/OPTICA.3.000222} {\bibfield  {journal}
  {\bibinfo  {journal} {Optica}\ }\textbf {\bibinfo {volume} {3}},\ \bibinfo
  {pages} {222} (\bibinfo {year} {2016})},\ \Eprint
  {http://arxiv.org/abs/1511.07339} {arXiv:1511.07339} \BibitemShut {NoStop}%
\end{thebibliography}%
\end{document}